\documentclass[aps,pra,twocolumn,superscriptaddress,amsmath,amssymb]{revtex4-1}
\usepackage{graphicx}
\usepackage{bm}
\usepackage{xcolor}
\usepackage{siunitx}

\newcommand{\tr}{\ensuremath{\operatorname{tr}}}

\newcommand{\sect}[1]{\par\textit{#1}.---\ignorespaces}
\newcommand{\SIrefs}{MacDonald1949a, Risken, Flindt2008a, Flindt2010a}
\newcommand{\supplement}{Supplemental Material \cite{supplement}\nocite{\SIrefs}}

\newcommand{\void}[1]{}

\begin{document}
	\title{Spectral properties of stochastic resonance in quantum transport}
	
	\author{Robert Hussein}
	\affiliation{Fachbereich Physik, Universit\"at Konstanz, D-78457 Konstanz, Germany}
	
	\author{Sigmund Kohler}
	\affiliation{Instituto de Ciencia de Materiales de Madrid, CSIC, E-28049 Madrid, Spain}
	
	\author{Johannes C. Bayer}
	\author{Timo Wagner}
	\author{Rolf J. Haug}
	\affiliation{Institut f\"ur Festk\"orperphysik, Leibniz Universit\"at Hannover, D-30167 Hanover, Germany}
	
	\date{\today}
	
	\begin{abstract}
		We investigate theoretically and experimentally stochastic resonance in a
		quantum dot coupled to electron source and drain via time-dependent tunnel
		barriers.  A central finding is a transition visible in the current noise
		spectrum as a bifurcation of a dip originally at zero frequency.  The
		transition occurs close to the stochastic resonance working point and relates to
		quantized pumping.  For the evaluation of power spectra from measured
		waiting times, we generalize a result from renewal theory to the ac driven
		case.  Moreover, we develop a master equation method to obtain
		phase-averaged current noise spectra for driven quantum transport.
	\end{abstract}
	
	\maketitle
	
	Stochastic resonance (SR) is a counter intuitive phenomenon by which the output
	signal of a device improves due to the action of external noise
	\cite{Jung1991a, Gammaitoni1998a}.  Typically it emerges as
	an interplay of periodic driving, nonlinearities, and noise-induced
	activation.  The paradigmatic example consists of two states separated by
	an energetic barrier, where an external oscillating force causes
	periodic transitions considered as signal.  When the force is rather weak,
	noise may help to cross the barrier and, thus, improves the signal.  For
	very strong noise, however, the output inherits too much randomness and
	degrades.  This reflects a prominent feature of SR, namely an optimal
	working point at an intermediate noise level.  A further characteristic
	property is that SR predominantly occurs when the driving frequency roughly
	matches one half the intrinsic decay rate of the system, $f=\Gamma_0/2$
	\cite{Gammaitoni1998a}.
	SR has been suggested as the mechanism behind very different phenomena
	ranging from the periodic recurrence of ice ages to biological signal
	processing \cite{Moss2004a}.  Many of these ideas have been realized 
	experimentally in the classical regime, while the quantum regime has been
	explored mainly theoretically \cite{Grifoni1996a, Wellens2000a, Lee2011a}.
	
	Recently in an experiment with a biased quantum dot with time-dependent
	tunnel rates, SR has been extended to the realm of quantum transport with
	the zero-frequency noise of the current as a measure for the signal quality
	\cite{Wagner2019a}.  It turned out that the current noise indeed assumes its
	minimum when the driving frequency obeys the mentioned SR condition.
	However, as only zero-frequency properties of the experimental data were
	evaluated, the question arises whether additional information can be
	extracted from the full power spectrum of the current fluctuations.
	
	With this letter, we demonstrate that the current noise spectrum
	provides relevant insight to the SR mechanism in quantum transport. We
	develop a method for computing the frequency-dependent Fano factor
	\cite{EmaryPRB2007a, Marcos2010a, HusseinPRB2014b} for ac driven
	transport making use of the time evolution of conditional cumulants
	\cite{Benito2016b} and compare the results with experimental data from an
	ac-driven quantum dot similar to Ref.~\cite{Wagner2019a}.  For the data
	analysis, we generalize the  relation between waiting times and the power
	spectrum of a spike train known from renewal theory \cite{Cox1962a,
	Gerstner2014a} to the ac-driven case.  Finally, we discuss possible
	applications for quantized charge pumping and current standards such as
	those of Refs.~\cite{Kaestner2015a, Platonov2015a}.
	
	\begin{figure}[b]
		\centerline{%
		\includegraphics[width=\columnwidth]{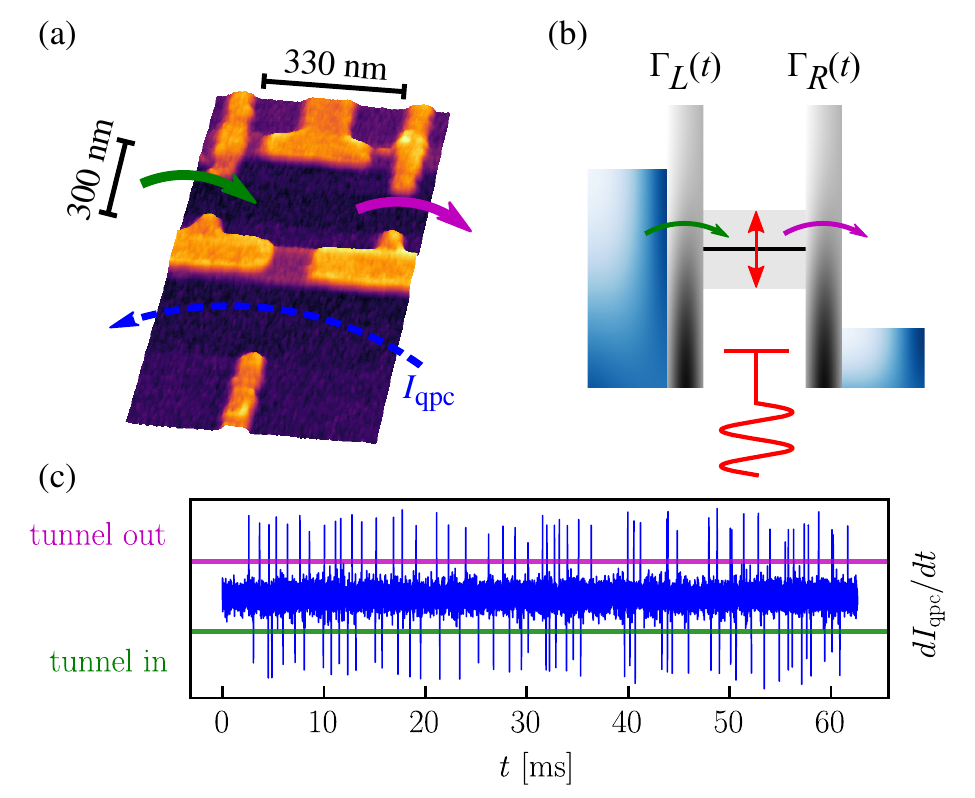}}
		\caption{(a) Strongly biased quantum dot with periodically time-dependent tunnel couplings. 
		It can be charged by electrons entering from the source (green arrow) and discharged
		when they leave to the drain (magenta).  The quantum point contact measures the dot occupation.
		(b) Corresponding theoretical model.
		(c) Time-derivative of $I_\text{qpc}$.  The sign of the spikes reflects
		the change of the dot occupation.}
		\label{fig:setup}
	\end{figure}
	
	\sect{Experimental setup and model}
	The experiments have been performed on a Schottky gate defined quantum dot
	based on the two-dimensional electron gas of a GaAs/AlGaAs heterostructure
	as shown in Fig.~\ref{fig:setup}(a).  An adjacent quantum point contact acts as charge
	monitor. By applying sufficiently negative voltages to the center gates,   
	the visible gap in the center is electrostatically closed and the two paths  
	are galvanically isolated. From the upper side, the quantum dot is confined
	by two tunnel barrier gates and a plunger gate, which are used to
	manipulate the tunneling rates and the energy levels of the quantum dot.   
	All measurements have been performed at \SI{1.5}{\kelvin}.
	
	The quantum dot is tunnel coupled to biased leads, where voltages are
	applied to the plunger gate and the tunnel barrier gates such that its
	lowest level, hosting up to one electron, lies in the center of the bias
	window. Care has been taken to tune the dot to a symmetric coupling to
	source and drain, such that the tunnel rates from the source and to the drain are equal.
	The ac components of the gate voltages let the dot level oscillate as
	sketched in Fig.~\ref{fig:setup}(b), and the tunnel rates becomes
	time-dependent \cite{Platonov2015a, Wagner2019a}.  We model this situation
	by the transition rates
	\begin{equation}
		\Gamma_{L/R}(t) = \Gamma_0 \exp[\pm\alpha_{L/R}A\cos(\Omega t)],
		\label{GammaLR}
	\end{equation}
	with the driving amplitude $A$ and a period $T=2\pi/\Omega \equiv 1/f$.  The
	leverage factors $\alpha_{L/R}$ as well as the intrinsic decay rate
	$\Gamma_0$ are adjusted such that the $\Gamma_{L/R}(t)$ match the rates in
	the experiment.  Transport phenomena in this open system
	can be described by a master equation of the form $\dot\rho =
	\mathcal{L}(t)\rho$, where $\rho$ is the reduced density operator of the
	central conductor with the $T$-periodic Liouvillian $\mathcal{L}(t) =
	\mathcal{L}(t+T)$, see the \supplement.
	
	Current measurements are affected by the
	displacement current of the fluctuating charge configurations
	\cite{Ramo1939a, Shockley1938a}. In our case, their origin is the
	stochastic charging and discharging of the quantum dot described by the
	jump operators, $\mathcal{J}_{L/R}$ and $\mathcal{J}_Q^\pm =
	\mathcal{J}_L^\pm - \mathcal{J}_R^\pm$ \cite{Marcos2010a, supplement},
	where $L$ and $R$ denote source and drain, respectively, while $Q$
	refers to the dot occupation.  As a consequence, current measurements in
	the leads of a mesoscopic two-terminal device provide the so-called total
	(or Ramo-Shockley) current $I  = -\kappa_L I_L+ \kappa_R
	I_R$, where $I_L$ and $I_R$ are the particle currents at the interfaces.
	$\kappa_L$ and $\kappa_R = 1-\kappa_L$ are normalized gate
	capacitances which we assume time independent and symmetric, $\kappa_L =
	\kappa_R = 1/2$.  While this distinction is irrelevant for the average
	current and the zero-frequency noise \cite{Kohler2005a}, it is quite
	important for the current noise spectrum \cite{Blanter2000a, Marcos2010a}.
	A measurement of the dot occupation via the charge
	monitor provides the full information about all currents.  Nevertheless, we
	focus on the total current, because it turns out that its noise spectrum is most
	significantly affected by SR.
	
	The relation between the noise of the total current $S_{\rm tot} $ and that of $I_L$,
	$I_R$, and $\dot Q$ is readily obtained from charge conservation,
	$N_L+N_R+Q = \text{const}$.  Hence, the corresponding current correlation
	function obeys $S_{\rm tot} = \kappa_L S_{L} + \kappa_R S_{R} - \kappa_L\kappa_R
	S_{Q}$, which holds in the time domain as well as in the
	frequency domain \cite{Blanter2000a, Marcos2010a}.
	
	\sect{Frequency-dependent Fano factor}
	We consider the particle current as the change of the electron number,
	given by the stochastic variable $j = \dot n$, in a region which may be a
	lead or the quantum dot (or any other compound of coupled quantum dots).
	Its symmetrized auto  correlation function $S(t,t') =
	\frac{1}{2}\langle[\Delta j(t),\Delta j(t')]_+\rangle$ in the stationary
	limit, must obey the discrete time-translation invariance of the
	Liouvillian, namely $S(t,t') = S(t+T,t'+T)$.  By introducing the time
	difference $\tau = t-t'$, one sees that $S(t,t-\tau)$ is invariant under
	$t\to t+T$, i.e., for constant $\tau$ it is $T$-periodic in $t$
	\cite{Kohler2005a}.  This implies that time-averages over a driving period
	are equivalent to averages over the phase of the driving \cite{Jung1990a}.
	Hence, we define the phase-averaged correlation function $\bar S(\tau)
	\equiv \overline{S(t+\tau,t)}^t = \overline{S(t,t-\tau)}^t$, where the
	second equality follows readily from simultaneous translation of all times.
	The corresponding phase-averaged spectral density $\bar S(\omega)$ is
	normalized to the average current $\bar I$ to yield as dimensionless noise
	spectrum the \textit{frequency-dependent Fano factor} $F(\omega) = \bar
	S(\omega)/\bar I$, which is our main quantity of interest.
	
	To compute $\bar S(\omega)$, we establish its relation to the conditional
	second moment of the electron number in the lead as $M_2(t|t') = \langle
	\Delta n^2(t)\rangle_{t'}$, where the subscript $t'$ denotes the reference
	time from which on we consider the fluctuations.
	Owing to $\dot n(t) = j(t)$, one finds \cite{supplement}
	\begin{equation}
		M_2(t|t_0) = \int_{t_0}^t dt''\int_{t_0}^t dt'\, S(t'',t') ,
	\end{equation}
	whose time derivative is the conditional second current cumulant $c_2(t|t_0)
	= 2\int_{t_0}^t dt' S(t,t')$.  Via the substitution $t' \to t'-\tau$,
	a subsequent phase average, and Fourier transformation, we obtain the
	generalized MacDonald formula \cite{supplement}  
	\begin{equation}
		\label{Sw}
		\bar S(\omega) = \omega\int_0^\infty d\tau\, \sin(\omega\tau)
		\int_0^T \frac{dt}{T} c_2(t+\tau|t) .
	\end{equation}
	The remaining tasks are the computation of the time evolution of the
	conditional current cumulant $c_2$ for sufficiently many values of $t$ (or initial
	phases) and a numerical $t$-integration.
	
	For this purpose, we employ a recent propagation method for the
	full-counting statistics \cite{Benito2016b}.  It is based on a generalized 
	master equation that consists of the Liouvillian and the jump operator of
	the considered current together with a counting variable
	\cite{Bagrets2003a}.  Taylor expansion in the counting variable provides
	the usual master equation $\dot\rho = \mathcal{L} \rho$ and the time
	dependent current expectation value, $I_\nu(t) = \tr (\mathcal{J}_\nu^+ -
	\mathcal{J}_\nu^-)\rho$ for $\nu=L,R,Q$.  The next order
	yields the second cumulant, $ c_2(t|t') =
	\tr(\mathcal{J}_\nu^++\mathcal{J}_\nu^-)\rho(t)
	+2\tr(\mathcal{J}_\nu^+-\mathcal{J}_\nu^-) X_1(t)$, where the auxiliary
	operator $X_1$ obeys the equation of motion $\dot X_1(t) = \mathcal{L}X_1 +
	[\mathcal{J}_\nu^+ -\mathcal{J}_\nu^- - I(t)]\rho(t) $.  Since we are
	interested in stationary correlations, the boundary condition at time $t'$
	is that (i) the density operator $\rho(t')$ must no longer contain
	transients and (ii) $X_1(t')=0$.
	For details of the derivation, see the \supplement.
	
	\sect{Power spectrum of the measured current}
	In the experiment, the charge monitor provides the times at which electrons
	tunnel from or to the dot, such that we can consider the current as a spike
	train as shown in Fig.~\ref{fig:setup}(c).  The experimental data at hand
	are the times between two subsequent tunnel events.  In the absence of the
	driving, the distribution function of these times relates to the power
	spectrum of the spike train
	\cite{Cox1962a, vanKampen1992a, BrandesAP2008a, Gerstner2014a}.  In the
	\supplement, we generalize this relation to the periodically time-dependent
	case and show that the phase-averaged power spectrum of the spike train
	reads  
	\begin{equation}
		\bar S(\tau) = \bar\gamma\delta(\tau) + \bar\gamma w(|\tau|) + \varphi(|\tau|) ,
		\label{S-wt}
	\end{equation}
	where the first term is the $\delta$-correlated shot noise for the mean
	spike rate $\bar\gamma$.  $w(\tau) = \sum_\ell w_\ell(\tau)$ is given by
	the probability distributions of the waiting times between a tunnel event
	and its $(\ell+1)$st successor, $w_\ell(\tau)$, which oscillate
	with the driving frequency \cite{Brange2020a} and which we sample
	from experimental data.
	
	For $\varphi$ we only know that it is $T$-periodic and has zero mean
	\cite{supplement}.  Without the driving, it vanishes such that
	Eq.~\eqref{S-wt} recovers a result from renewal theory~\cite{Cox1962a,
	Gerstner2014a}.  To determine $\varphi$, we notice that
	for large time difference $\tau$, the tunnel events are uncorrelated and,
	thus, $\bar S$ vanishes.  Therefore, $\varphi$ can be identified as the
	long-time oscillations of $w(\tau)$.  Accordingly in the
	frequency domain we use the fact that finite-time Fourier transformation
	converts long-time oscillations to poles of first order, while $\bar
	S(\omega)$ is expected to be a smooth function.  Therefore, poles in the
	Fourier transformed of $w(\tau)$ can be attributed to $\varphi$.  They can
	be determined by fitting.
	
	\sect{SR signatures in the Fano factor}
	\begin{figure}
		\centerline{%
		\includegraphics{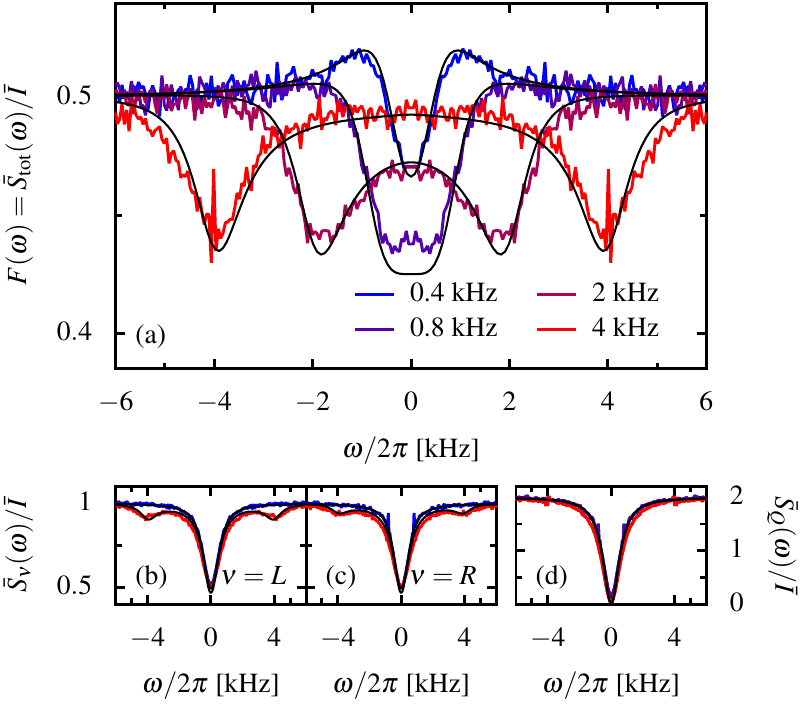}}
		\caption{%
		Frequency-dependent Fano factor, i.e., normalized power spectrum of the
		total current (a), the current at source (b), drain (c), and the net
		current to the dot (d)
		for driving amplitude $A=\SI{10}{\meV}$ and various driving frequencies.
		The colored lines mark experimental data, while the black lines are
		computed with the master equation approach.  The other parameters are
		$\alpha_L=0.09$, $\alpha_R=0.065$, and $\Gamma_0=\SI{1.675}{\kHz}$.
		}
		\label{fig:ExpTheo}
	\end{figure}%
	In Ref.~\cite{Wagner2019a}, the existence of SR in quantum transport has
	been demonstrated with the zero-frequency Fano factor as a noise measure.
	However, a complete picture of the noise must include its full spectral
	properties.  As a reference, let us first mention that in the absence of
	driving, the current in a symmetric quantum dot has white noise
	characterized by the constant Fano factor $F(\omega)=1/2$
	\cite{Korotkov1994a, Gustavsson2006a, Ubbelohde2012a}.  Moreover, for adiabatic driving, the
	symmetry gets lost such that most of the time the zero-frequency noise is
	enhanced \cite{RiwarPRB2013a}.  Since we are interested in SR, we consider
	much larger frequencies of the order $\Gamma_0$.
	Figure~\ref{fig:ExpTheo}(a) shows noise spectra of the total current for
	various nonadiabatic driving frequencies.  For the relatively low frequency
	$f=\SI{0.4}{\kHz}$, the zero frequency noise is already smaller than the
	standard value $1/2$ expected for the undriven dot.  In addition, in the vicinity of
	$\omega=0$, however, $F(\omega) > 1/2$.  With increasing frequency $f$, we
	witness the dip in the noise spectrum at $\omega=0$ becoming deeper and
	broader.  Close to the SR condition $f\approx\Gamma_0/2$, the dip evolves
	into a double dip located at the driving frequency $\pm f$
	($f=\SI{2}{\kHz}$ and $\SI{4}{\kHz}$), which underlines the importance of
	considering the whole noise spectrum.  While the zero-frequency noise
	insinuates disappearance of the SR effect---and becomes almost insensitive
	to it at $f=\SI{4}{\kHz}$---the frequency-dependent analysis reveals that
	the noise suppression remains, but occurs in the spectrum at finite
	frequency.  In contrast, the current noise at source and drain depends only
	weakly on the driving, as can be seen in Figs.~\ref{fig:ExpTheo}(b) and
	\ref{fig:ExpTheo}(c).  Only when the driving frequency exceeds the SR
	frequency, i.e., for $f\gtrsim \Gamma_0/2$, the Fano factor of the source and
	drain currents develop small dips at $\omega \approx \pm 2\pi f$.
	Interestingly, the
	noise of the dot current [Fig.~\ref{fig:ExpTheo}(d)] and, thus, that of the
	dot occupation are practically independent of the driving.  This emphasizes
	that for transport SR, the noise properties are primarily manifest in the
	total current.
	
	\begin{figure}
		\centerline{%
		\includegraphics{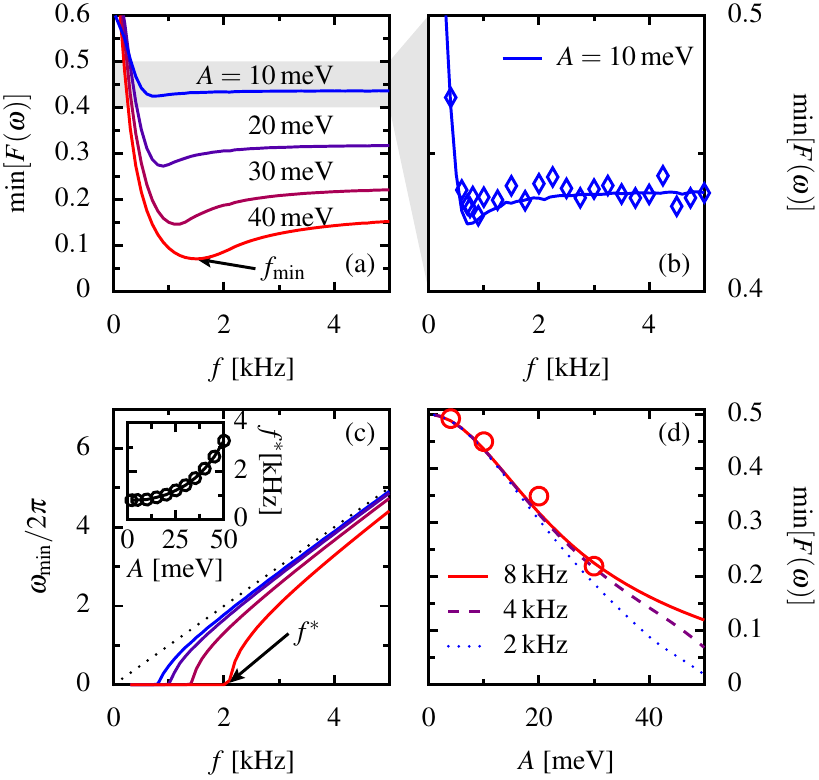}}
		\caption{Analysis of the dips in the noise spectra.
		(a) Minimum of the Fano factor $F(\omega)$ for various amplitudes as
		function of the driving frequency.
		(b) Enlargement of the shaded area in panel (a) together with
		corresponding experimental data indicated by diamonds.
		(c) Frequency at which the minimum is located in the power spectrum for the
		data in panel (a).
		Inset: transition frequency $f^*$ as a function of the driving amplitude.
		(d) Minimum of the Fano factor as a function of the driving amplitude for
		various driving frequencies.  The circles mark experimental results for
		$f=\SI{8}{\kHz}$.
		All other parameters are as in Fig.~\ref{fig:ExpTheo}.
		}
		\label{fig:Fmin}
	\end{figure}%
	The magnitude and the position of the noise reduction is analyzed in
	Fig.~\ref{fig:Fmin}.  Panels (a) and (b) show how the minimum of
	$F(\omega)$, as observed in Fig.~\ref{fig:ExpTheo}(a),  changes with the
	driving frequency $f$. The data confirm that the minimum of the Fano factor
	in the adiabatic limit, i.e.\ low driving frequencies, assumes values
	considerably larger than the standard value $1/2$, as discussed above.
	Upon increasing the driving frequency, the minimum becomes lower until at
	an amplitude-dependent value $f_\mathrm{min}$, it starts to increase again.
	Figure~\ref{fig:Fmin}(b) shows a nice agreement between theory and
	experiment for the development of the minimum of the Fano factor for an
	amplitude $A=\SI{10}{\meV}$ considering a certain noise in the experimental
	data. In particular, the data clearly confirm the frequency independence of
	the minimum beyond the SR point.
	Next we consider the location of the minimum in the spectrum,
	$\omega_\mathrm{min}$, depicted in Fig.~\ref{fig:Fmin}(c), where the
	transition from $\omega_\mathrm{min}=0$ to a finite value corresponds to
	the splitting of the dip as observed in Fig.~\ref{fig:ExpTheo}(a).  This
	happens at a frequency $f^*$ which increases with increasing driving
	amplitude. In the inset of Fig.~\ref{fig:Fmin}(c)  the detailed dependence
	on the driving amplitude is plotted.  The growth of $f^*$ with the driving
	amplitude reminds one to the shift of the working point which has also been
	observed for the ``usual'' SR in closed systems \cite{Gammaitoni1998a}.
	
	Figure~\ref{fig:Fmin}(d) depicts the minimum of the Fano factor as function
	of the driving amplitude for three driving frequencies (beyond the SR). A
	strong reduction of the Fano factor is clearly observed, as also seen in
	the experimental data (marked by circles) for a driving frequency of
	$\SI{8}{\kHz}$ \footnote{As the experimental spectra are noisy, we have convoluted
	them with a Gaussian of width $\sigma=\SI{100}{\Hz}$ before reading off the
	minima.}.  At strong driving the reduction is enhanced for frequencies
	closer to the SR condition, see the curve for $\SI{2}{\kHz}$ in comparison
	with the curve at $\SI{8}{\kHz}$.
	
	\begin{figure}
		\centerline{%
		\includegraphics{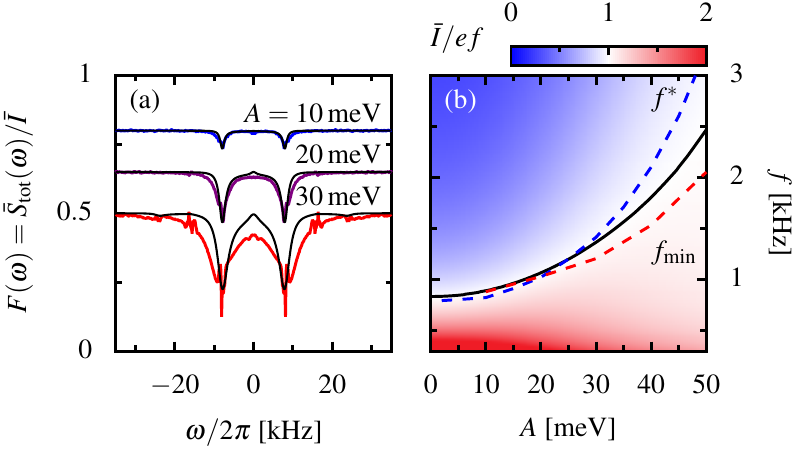}}
		\caption{(a) Frequency-dependent Fano factor of the total current for
		driving frequency $f=\SI{8}{\kHz}$ and various amplitudes.
		All other parameters are as in Fig.~\ref{fig:ExpTheo}.
		The curves for \SI{10}{\meV} and \SI{20}{\meV} are vertically shifted by
		0.15 and 0.3, respectively.
		(b) Transported charge per driving period, where the solid line highlights
		parameters with quantized current $\bar I=ef$.  The dashed lines
		mark the transition frequency $f^*$ as a function of the amplitude
		[see inset of Fig.~\ref{fig:Fmin}(c)] and the frequency $f_\text{min}$ at
		which for given $A$ the Fano factor assumes its minimum
		[see Fig.~\ref{fig:Fmin}(a)].}
		\label{fig:pumping}
	\end{figure}%
	
	To investigate the amplitude dependence in more detail
	Fig.~\ref{fig:pumping}(a) shows the frequency dependent Fano factor at
	three different applied amplitudes for a driving frequency much larger than
	$f^*$.  The curves exhibit the typical double-dip structure discussed
	above.  With an increasing amplitude, the shape of the Fano factor starts
	to deviate from the Lorentzian obtained for weak driving.  Moreover, for
	$A=\SI{30}{\meV}$, we witness that the impact of non-linearities is visible
	as a tiny additional dip at $\omega/2\pi \approx 3f$.  In the experimental
	data the additional dip is less clear, because the large driving amplitude
	makes it increasingly difficult to determine with sufficient precision the
	poles stemming from the last term in Eq.~\eqref{S-wt} of
	Ref.~\cite{supplement}. The increased broadening of the dips with
	increasing amplitude is, by contrast, even more expressed in the
	experimental data.
	
	Recent interest in controlled single-electron tunneling stems from the
	challenge of building current standards \cite{Kaestner2015a} that transport
	an integer number of electrons per cycle.  Let us therefore discuss the
	frequency-dependent Fano factor in this context.
	Figure~\ref{fig:pumping}(b) shows the average (electric) current as a
	function of the driving amplitude and frequency. In the same figure also
	the current $\bar I = e f$ (black line) and the above discussed frequencies
	$f_\mathrm{min}$ (red dashed line) and $f^*$ (blue dashed line) are shown
	as a function of the driving amplitude. For amplitudes smaller than $A
	\approx \SI{25}{\meV}$, the three lines more or less overlap, i.e.\ there is
	no clear difference between $f_\mathrm{min}$ and $f^*$ and they mark the
	quantized current with $\bar I = e f$. For larger amplitudes a plateau-like
	structure with current $\bar I\approx ef$ is observed [white region in
	Fig.~\ref{fig:pumping}(b)] as expected from the experiments investigating
	single-electron pumping for current standards \cite{Kaestner2015a}. The
	line on which $\bar I = e f$ is fulfilled exactly lies in the middle of
	this plateau and between transition frequency $f^*$ and the frequency
	$f_\mathrm{min}$ at which the Fano factor is minimal. 
	With increasing amplitude, both the
	width of the plateau and the difference between $f_\mathrm{min}$ and $f^*$
	become larger.  Accordingly, for large frequencies, very low Fano factors
	require larger amplitudes, see Fig.~\ref{fig:Fmin}(d).  Such a plateau
	widening with increasing amplitude was also observed in the pumping
	experiments investigating current standards, see e.g.\
	Refs.~\cite{Blumenthal2007a, Kaestner2008a}. Interestingly, the two frequencies
	$f_\mathrm{min}$ and $f^*$ mark the borders of the plateau. In this way an
	analysis of the frequency dependent Fano factor can help to optimize the
	pumping conditions for the current standard.
	
	\sect{Conclusions}
	We have analyzed experimentally and theoretically the frequency dependent
	current noise in a transport SR experiment.  The most noticeable effect is
	visible in the power spectrum of the total current as a splitting of a dip
	at zero frequency to a double dip located at the driving frequency.  For
	small amplitudes, the transition between these two qualitatively different
	regimes occurs when the SR condition is met.  With increasing amplitudes,
	the transition frequency shifts towards larger values. Our results show the
	relation between transport SR and quantized electron pumping used for
	current standards.
	
	\begin{acknowledgments}
		This work was supported by the Zukunftskolleg of the University of Konstanz
		and by the Spanish Ministry of Science, Innovation, and Universities under
		grant No.\ MAT2017-86717-P, by the Deutsche Forschungsgemeinschaft (DFG,
		German Research Foundation) under Germnany's Excellence Strategy -- EXC-2123
		Quantum Frontiers -- 390837967, and by the State of Lower Saxony, Germany,
		via Hannover School for Nanotechnology and School for Contacts in Nanosystems.
	\end{acknowledgments}
	
	\section*{Supplemental Material}
	\appendix
	\renewcommand{\thefigure}{S\arabic{figure}}
	\setcounter{figure}{0}
	
	\section{Master equation and jump operators}
	\label{app:ME}
	
	We consider a quantum dot that for energetic reasons can be occupied with
	at most one electron, such that the dynamics is restricted to the
	spin states $|{\uparrow}\rangle$, $|{\downarrow}\rangle$, and the
	empty state $|0\rangle$ with probabilities $P_\uparrow$, $P_\downarrow$,
	and $P_0$.  For weak tunnel coupling and large bias, electrons can enter
	only from the left lead (source) and will leave to the right lead (drain).
	In the absence of magnetic fields and spin effects, this situation is
	captured by the Markovian master equation
	\begin{equation}
		\frac{d}{dt}
		\begin{pmatrix}
			P_\uparrow\\P_\downarrow \\ P_0
		\end{pmatrix} 
		=
		\begin{pmatrix}
			-\Gamma_R & 0 & \Gamma_L/2 \\
			0 & -\Gamma_R & \Gamma_L/2 \\
			\Gamma_R & \Gamma_R & -\Gamma_L
		\end{pmatrix}
		\begin{pmatrix}
			P_\uparrow\\P_\downarrow \\ P_0
		\end{pmatrix},
	\end{equation}
	where the factor 1/2 for the source-dot tunneling is introduced for the ease of notation. 
	In our case, the rates $\Gamma_{L/R}$ are periodically time-dependent owing
	to an ac gate voltage applied to the tunnel barriers.
	Introducing the probability $P_1 = P_\downarrow + P_\uparrow$ for the occupation 
	with any spin projection, we obtain a master equation 
	$\dot\rho = \mathcal{L}\rho$ for $\rho =(P_1,P_0)$ with the Liouvillian
	\begin{equation}
		\mathcal{L} =
		\begin{pmatrix}			-\Gamma_R & \Gamma_L \\
		\Gamma_R & -\Gamma_L \end{pmatrix}		.
	\end{equation}
	Using the convention that an upper index $+$ refers to currents flowing from
	the central conductor of one of the leads, the corresponding jump operators
	read \cite{Bagrets2003a}
	\begin{equation}
		\mathcal{J}_{L}^- =
		\begin{pmatrix}		0 & \Gamma_L \\ 0 & 0 \end{pmatrix}		,
		\quad
		\mathcal{J}_{R}^+ =
		\begin{pmatrix}		0 & 0 \\ \Gamma_R & 0 \end{pmatrix}		,
	\end{equation}
	while for uni-directional transport, $\mathcal{J}_{L}^+ = \mathcal{J}_{R}^-
	= 0$.
	
	\section{Frequency-dependent Fano factor}
	\label{app:formalism}
	
	We consider a Markovian master equation $\dot\rho = \mathcal{L}(t)\rho$ for
	the reduced density operator of the conductor with a periodically time
	dependent Liouvillian $\mathcal{L}(t) = \mathcal{L}(t+T)$.  The
	time-dependence may affect the conductor itself as well as the leads or the
	conductor-lead couplings.  Our goal is a propagation method for the
	computation of the stationary symmetric current-current correlation function
	\begin{equation}
		S(t,t') = \frac{1}{2}\langle[\Delta j(t),\Delta j(t')]_+\rangle ,
		\label{app:Stt}
	\end{equation}
	where $j$ is defined as the time derivative of the particle number $n$ in a
	given region such as the leads or the central conductor, while $\Delta j =
	j-\langle j\rangle$.
	
	An alternative characterization of current fluctuations is the full
	counting statistics of the transported particles with the conditional
	moments
	\begin{equation}
		M_k(t|t') = \langle n^k(t)\rangle_{t'}
	\end{equation}
	with the boundary condition $M_k(t'|t')=0$ for all $k>0$.
	Both concepts are related by the MacDonald formula \cite{MacDonald1949a}
	which for \textit{time-independent} problems establishes a connection
	between the current correlation function \eqref{app:Stt} and the variance
	of the number of transported particles.  In a first step, we generalize
	this formula to the \textit{time-dependent} case and subsequently use the
	result as a basis for computing the phase-averaged current noise spectrum.
	
	\subsection{MacDonald formula for time-dependent transport}
	
	The moments $M_k$ of a probability distribution contain the same
	information as the corresponding cumulants (or irreducible moments)
	\cite{Risken}.  For a Markovian transport process, the latter eventually
	grow linear in time, which motivates the definition of current cumulants as
	their time derivatives \cite{Bagrets2003a}.  Obviously, the first current
	cumulant, i.e., the rate by which the number of transported particles grows
	is the current, while the second current cumulant is the time derivative of
	the variance, $c_2(t|t') = \frac{d}{dt}[ M_2(t|t') - M_1^2(t|t')]$.
	Upon noticing that $\dot n = j$, the definitions of $S$ and $c_2$
	directly provide the relation
	\begin{equation}
		\label{app:CS}
		c_2(t|t') = 2 \int_{t'}^t dt''\, S(t,t'') .
	\end{equation}
	
	For time-independent systems, two-time expectation values such as $S(t,t')$
	are called stationary if they are homogeneous in time, i.e., if $S(t,t') =
	S(t+\tau,t'+\tau)$ for any time translation $\tau$.  Setting $\tau=-t'$
	yields the known fact that stationary correlation functions depend only on
	time differences, $S(t,t') = S(t-t')$.  Then Eq.~\eqref{app:CS} in Fourier
	representation reads $c_2(\omega) = 2iS(\omega)/\omega$, which allows one
	to compute $S(\omega)$ directly from the time evolution of $c_2$
	\cite{MacDonald1949a}.
	
	For periodically driven systems, time translation invariance is granted
	only for time shifts by multiples of the driving period $T$.  Hence,
	stationarity corresponds to the weaker relation
	\begin{equation}
		\label{app:stat}
		S(t,t') = S(t+T, t'+T) .
	\end{equation}
	Nevertheless, one can define a noise measure that depends on only a single time
	argument, namely the phase-averaged correlation function
	\cite{Gammaitoni1998a}, which is equivalent to the average over one driving
	period when keeping the time difference $\tau = t-t'$ constant.  Hence, we
	can characterize the current fluctuations by the function
	\begin{equation}
		\label{app:Stau}
		\bar S(\tau) \equiv \int\limits_\text{period} \frac{dt}{T} S(t+\tau,t) .
	\end{equation}
	As the integrand in this expression is periodic in $t$ and by definition
	symmetric in its time arguments, one can easily see that $\bar S(\tau)=\bar
	S(-\tau)$.
	
	To find a relation between the conditional current cumulant
	and the correlation function, we substitute in Eq.~\eqref{app:CS} the
	primed times by the time differences $\tau=t-t'$ and $\tau'=t-t''$ and
	introduce the time shift $t\to t+\tau$ to obtain
	\begin{equation}
		\label{app:CS2}
		c_2(t+\tau|t) = 2 \int_{0}^{\tau} d\tau'\, S(t+\tau,t+\tau-\tau') .
	\end{equation}
	According to Eq.~\eqref{app:stat}, the right-hand side of this relation is
	$T$-periodic in $t$, and hence the left-hand side as well.  Therefore, we
	can perform an average over the driving period and after taking the
	derivative with respect to $\tau$, we obtain
	\begin{equation}
		\bar S(\tau) = \frac{1}{2} \frac{d}{d\tau} \bar c_2(\tau)
		\label{app:MDt}
	\end{equation}
	with the time-averaged second current cumulant
	\begin{equation}
		\bar c_2(\tau) = \int \frac{dt}{T} c_2(t+\tau|t) .
		\label{app:C2}
	\end{equation}
	We will see below that $c_2$ can be computed by numerical propagation of a
	generalized master equation.
	
	In a final step, we use the symmetry $\bar S(\tau) = \bar S(-\tau)$ to
	write $\bar S(\omega)$ as a Fourier cosine transformed and integrate by parts
	to obtain
	\begin{equation}
		\label{app:MDw}
		\bar S(\omega) = \omega\int_0^\infty d\tau\, \sin(\omega\tau)
		\bar c_2(\tau) ,
	\end{equation}
	which represents the requested generalization of the MacDonald formula for
	periodically driven conductors.  It relates the phase-averaged noise
	spectrum $\bar S(\omega)$ to the conditional second cumulant.
	For time-independent problems, $c_2(t+\tau|t) = c_2(\tau)$, such
	that the $t$-integration in Eq.~\eqref{app:C2} becomes trivial and one
	recovers the classic MacDonald formula \cite{MacDonald1949a}.
	
	Let us remark that the zero-frequency limit of Eq.~\eqref{app:MDw} obeys
	the relation
	\begin{equation}
		\lim_{\omega\to 0} \bar S(\omega) = \lim_{\tau\to\infty} c_2(t+\tau|t),
		\label{app:S0}
	\end{equation}
	which is independent of $t$.  For periodically driven conductors, this may
	be proven upon noticing that the right-hand side of Eq.~\eqref{app:CS} is
	the zero-frequency limit of a Fourier integral \cite{Kohler2005a} or via
	the long-time theorem of Laplace transformation.
	
	\subsection{Propagation method for cumulants}
	
	The time evolution of the current cumulants for a time-dependent transport
	problem can be computed with the hierarchical scheme derived in
	Ref.~\cite{Benito2016b}.  Here we briefly sketch the underlying ideas and
	adapt them to our needs.
	
	We consider the number of electrons $n$ in one of the leads as the main
	observable.  To keep track of this degree of freedom even after tracing it
	out, we introduce a counting variable $\chi$ and define the moment
	generating function $Z(\chi) = \langle e^{i\chi n}\rangle$.  For its
	computation, we employ a generalized reduced density operator $R(\chi)$
	constructed such that $\tr R(\chi) = Z(\chi)$.  It obeys the master
	equation (we omit obvious time arguments) \cite{Bagrets2003a}
	\begin{align}
\label{app:dotR}
\frac{d}{dt} R(\chi) ={}& [\mathcal{L} + \mathcal{J}(\chi)] R(\chi) , \\
\mathcal{J}(\chi) ={}& (e^{i\chi}-1)\mathcal{J}^+ + (e^{-i\chi}-1)\mathcal{J}^- ,
	\end{align}
	where the jump operators $\mathcal{J^\pm} = \mathcal{J}_{L/R}^\pm$ describe
	incoherent electron tunneling from the conductor to the respective lead and
	back.
	
	The associated current cumulants are the derivatives of the generating
	function $\phi(\chi) = \frac{d}{dt}\ln Z(\chi)$ and read $c_k =
	(\partial/\partial{i\chi})^k \phi(\chi)|_{\chi=0}$ \cite{Bagrets2003a}.
	Defining $X(\chi) = R(\chi)/Z(\chi)$, Taylor expansion of the
	generalized master equation \eqref{app:dotR}, $\phi(\chi)$, and $X(\chi)$
	yields the iteration \cite{Benito2016b}
	\begin{align}
\label{app:Ck}
		c_k ={}& \sum_{k'=0}^{k-1} \begin{pmatrix}		k\\k'\end{pmatrix}
\tr\mathcal{J}^{(k-k')} X_{k'} ,
\\
\label{app:Xk}
\dot X_k ={}& \mathcal{L}X_{k}
		+ \sum_{k'=0}^{k-1} \begin{pmatrix}		k\\k'\end{pmatrix}
  \{\mathcal{J}^{(k-k')}-c_{k-k'}\}X_{k'} ,
	\end{align}
	where
	\begin{equation}
		\mathcal{J}^{(k)} = \mathcal{J}^+ + (-1)^k\mathcal{J}^-
		\label{app:Jk}
	\end{equation}
	while $c_k$ and $X_k$ are the Taylor coefficients of $\phi$ and $X$,
	respectively.
	Notice that for $k=0$, Eq.~\eqref{app:Xk} is the usual master equation for
	the reduced density operator $X_0$.  For \textit{time-independent}
	problems, this scheme is equivalent to the iteration derived in
	Refs.~\cite{Flindt2008a, Flindt2010a}.  Truncating Eqs.~\eqref{app:Ck} and
	\eqref{app:Xk} at second order provides the time dependent current
	expectation value and the corresponding second cumulant.
	
	The initial condition of the differential equation \eqref{app:Xk} deserves
	some attention.  Let us recall that we are interested in the stationary
	conditional moments, which means that at initial time $t_0$ of the
	integration, (i) all moments and cumulants must vanish and (ii) transients
	of the density operator must have decayed already.  Condition (i) is
	granted when $X(\chi)$ at $t_0$ is $\chi$-independent.  Thus, all its
	Taylor coefficients vanish with the exception of $X_0$.  For
	$X_0$, condition (ii) means that it must be the steady-state density
	operator at $t_0$, $X_0(t_0) = \rho_\infty(t_0)$.  It is computed by
	starting the propagation at some time $t_0-t_\text{trans}$, where
	$t_\text{trans}$ is the timescale on which transients decay.

	\subsection{Ramo-Shockley theorem}
	
	Currents in the leads are affected by the displacement
	current of the electric field of the fluctuating charge distribution.
	While this is usually irrelevant for the average current, it may play a
	significant role for the current noise at finite frequency
	\cite{Blanter2000a}.  Therefore, it is here important to notice that the
	measured current in the leads (also referred to as the total current)
	flowing from the left to the right lead (with electron numbers $N_{L/R}$) 
	is a weighted average of the currents at both contacts. Using the sign 
	convention $I_{L/R} = \dot N_{L/R}$, it is given by 
	\begin{equation}
		\label{app:IRS}
		I_\mathrm{tot} = -\kappa_L I_L + \kappa_R I_R ,
	\end{equation}
	where the weights $\kappa_L$ and $\kappa_R$ reflect the capacitances at the
	contacts normalized such that $\kappa_L+\kappa_R=1$. We assume that the driving
	does not affect these capacitances.

	As the total charge is conserved, the number of electrons on the conductor
	changes as $\dot Q = -I_L -I_R$.  Then the correlation $S_\mathrm{tot}$ of the total
	current becomes
	\begin{equation}
		S_\mathrm{tot} = \kappa_L S_{L} + \kappa_R S_{R} - \kappa_L\kappa_R S_{Q} ,
	\end{equation}
	which holds in the time domain and in the frequency domain, as well as for
	the phase average in Eq.~\eqref{app:Stau}.
	While $S_{L}$ and $S_{R}$ can be computed as described above, the direct
	computation of $S_\mathrm{tot}$ and $S_{Q}$ needs more care, because
	generally, one cannot simply add the jump operators of the left and the
	right lead as insinuated by Eq.~\eqref{app:IRS}.
	
	To derive the jump operators for the total current and for the displacement
	current, we employ ideas of Ref.~\cite{Marcos2010a}.  We start
	from the integrated form of Eq.~\eqref{app:IRS} and charge conservation,
	i.e.\ $N_\mathrm{tot} = -\kappa_L N_L + \kappa_R N_R$ and $Q = -N_L -N_R$, respectively,
	where $N_\mathrm{tot}$ is the number of electrons transported with the total current.
	Then we write the exponent in the expression for the moment generating
	function $Z(\chi_L,\chi_R)$, in terms of $N$ and $Q$ as
	\begin{equation}
		\chi_L N_L + \chi_R N_R 
		= (-\kappa_R\chi_L-\kappa_L\chi_R)Q +(\chi_R-\chi_L)N_\mathrm{tot},
		\nonumber
	\end{equation}
	which lets us conclude that the counting variables for $N_\mathrm{tot}$ and $Q$ read
	$\chi_{\rm tot} =\chi_R -\chi_L$ and $\chi_Q = -\kappa_R\chi_L - \kappa_L\chi_R$,
	respectively.  In turn,
	\begin{align}
		\begin{split}
\chi_L ={}& {-}\kappa_L\chi_{\rm tot} - \chi_Q ,
\\
\chi_R ={}& \phantom{-} \kappa_R\chi_{\rm tot} - \chi_Q .
		\end{split}
	\end{align}
	Now the corresponding jump operators follow simply by differentiation of
	the Liouvillian with respect to the associated counting variable using the
	chain rule \cite{Marcos2010a},
	\begin{align}
		\begin{split}
\mathcal{J}_{\rm tot}^{(k)}
={}& \partial_{i\chi_{\rm tot}}^k
\mathcal{L}(\chi_L,\chi_R)\big|_0
= (-\kappa_L)^k \mathcal{J}_L^{(k)} + \kappa_R^k \mathcal{J}_R^{(k)} ,
\\
\mathcal{J}_Q^{(k)}
={}& \partial_{i\chi_Q}^k
\mathcal{L}(\chi_L,\chi_R)\big|_0
= (-1)^k \mathcal{J}_L^{(k)} + (-1)^k \mathcal{J}_R^{(k)} .
		\end{split}
\label{app:JTotQ}
	\end{align}
	In the final expression no mixed derivatives occur, because the contribution of each lead
	enters as a separate term.  Interestingly, the first-order jump operators,
	$\mathcal{J}_{\rm tot}^{(1)}$ and $\mathcal{J}_Q^{(1)}$, are as naively expected,
	while in particular the higher orders of $\mathcal{J}_{\rm tot}^{(k)}$ contain powers
	of the contact capacities.  Notice that a possibly different sign
	convention for the currents merely leads to a global prefactor $(-1)^k$
	which is irrelevant for derivatives of even order and, thus, for the
	current correlation function.
	
	\subsection{Numerical scheme}
	For the practical computation, we start from the Liouvillian for the
	conductor, identify the jump terms and attribute them counting variables to
	find $\mathcal{J}_{L/R}$ and, thus, $\mathcal{J}_{\rm tot}$ and
	$\mathcal{J}_Q$.  Then we use the iteration in Eqs.~\eqref{app:Ck} and \eqref{app:Xk}
	to obtain for each current the corresponding conditional centered second moment
	$c_2(t+\tau|t)$ as function of $\tau$ for different initial times $t$.
	Averaging over $t$ yields $\bar c_2(\tau)$, which we finally insert into
	Eq.~\eqref{app:MDw}.  The remaining Fourier integral is conveniently
	evaluated via discrete Fourier transformation.  Since the zero-frequency
	contribution of the resulting spectrum may be rather sensitive to numerical
	noise, we determine it from the long-time limit in Eq.~\eqref{app:S0}.

	\section{\label{app:renewal}Current noise from waiting-time distributions}
	We consider the current as a spike train
	\begin{equation}
		j(t) = \sum_k \delta(t-s_k)
		\label{app:spiketrain}
	\end{equation}
	with tunnel events at times $s_k$, which in the experiment can be obtained by
	monitoring the dot occupation.  However, the $s_k$ may not be known with
	sufficient precision to exclude effects of possible long-time phase drifts of the external
	driving.  Therefore, we will base the evaluation on the differences between
	subsequent events, $t_k = s_k-s_{k-1}$, which also provide the times
	between events with $\ell$ events in between.  In practice, each data set
	consists of the order $10^6$ events, while for $\ell$ values up to several
	$\sim 100$ are sufficient. Our goal is now to obtain from these data the phase-averaged
	power spectral density $\bar S(\tau)$ of the spike train.  For this purpose, we generalize a
	relation between the power spectrum and the waiting time statistics known
	from renewal theory \cite{Cox1962a, Gerstner2014a} to periodically time-dependent systems.
	
	The power spectrum of a (time-independent) renewal processes can be
	computed from the waiting time distribution between two subsequent events,
	$w_0(\tau)$.  The derivation makes use of the fact that the waiting time
	distributions $w_\ell(\tau)$ of the $(\ell+1)$st successor of an event,
	$w_\ell(\tau)$ for $\ell>0$ can be computed simply by an $\ell$-fold auto
	convolution of $w_0$ \cite{Cox1962a, Gerstner2014a}.  However, this relation holds true only in the
	\textit{time-independent} case, while we will sample the
	$w_\ell(\tau)$ from our experimental data.  In the following, we
	demonstrate that in periodically time-dependent cases, knowledge
	of the $w_\ell$ allows one to compute the \textit{phase-averaged} power
	spectrum $\bar S(\omega)$.
	
	Let us start by noticing that for the spike train in
	Eq.~\eqref{app:spiketrain}, any contribution to the expectation value
	$\langle j(t+\tau)j(t)\rangle$ originates from spikes at times $t$ and
	$t+\tau$, where for the moment we assume $\tau>0$.  Therefore, this
	expectation value must be proportional to the joint probability density for
	spikes at times $t$ and $t+\tau$, i.e.,
	\begin{equation}
		\langle j(t+\tau)j(t)\rangle = \lambda p(t+\tau,t) ,
	\end{equation}
	where $p(t',t)dt\,dt'$ is the probability for an event in the time interval
	$[t,t+dt)$ and a further one in the \textit{different} interval
	$[t',t'+dt')$.  The proportionality factor $\lambda$ will be determined
	later.  Subtracting from the left-hand side the product of the expectation
	values $\gamma(t) = \langle j(t)\rangle$ and $\gamma(t+\tau)$ obviously yields the auto
	correlation of the spike train,
	\begin{equation}
		S(t+\tau,t) = \langle j(t+\tau)j(t)\rangle - \gamma(t+\tau)\gamma(t) .
	\end{equation}
	Owing to the periodic time-dependence of $\gamma(t)$, the $t$-average of
	the last term is not only the square of the mean spike rate, $\bar\gamma
	=\overline{\gamma(t)}^t$, but in addition contains a $T$-periodic term
	$\varphi(\tau)$ with zero mean.  Thus, the phase-averaged auto correlation
	becomes
	\begin{equation}
		\bar S(\tau) = \lambda\overline{p(t+\tau,t)}^t - \bar\gamma^2 - \varphi(\tau) .
		\label{app:Sbasic}
	\end{equation}
	
\begin{figure}
		\centerline{%
		\includegraphics{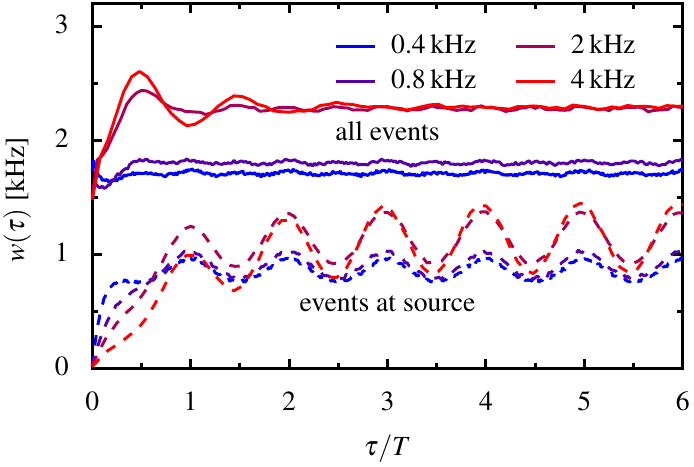}}
		\caption{Summed waiting time distributions $w(\tau) = \sum_\ell
		w_\ell(\tau)$ sampled from the experimental data for the driving frequency
		$f=\SI{2}{\kHz}$, where $\tau$ is given in units of the driving period $T=1/f$.
		Solid lines mark the result for all events, while dashed lines
		consider only events at the source.  The corresponding result
		for the drain (not shown) looks very similar to the latter.  All other
		parameters are as in Fig.~\ref{fig:ExpTheo} of the main text.}
		\label{fig:wtau}
	\end{figure}

	Next we relate the first term on the right-hand side of
	Eq.~\eqref{app:Sbasic} to the waiting-time distributions.  To this end, we
	notice that $w_\ell(\tau)$ is the $t$-averaged probability density for
	finding a spike at time $t+\tau$ conditioned to the occurrence of a spike at
	time $t$ and $\ell$ spikes in between.  Thus,
	\begin{equation}
		w_\ell(\tau) = \langle p_\ell(t+\tau|t)\rangle_t
		= \overline{p_\ell(t+\tau|t) p(t)}^t ,
		\label{app:wtau}
	\end{equation}
	where $\langle x(t)\rangle_t = \frac{1}{T}\int_0^T dt\, x(t) p(t) =
	\overline{x(t)p(t)}^t$ denotes the expectation value of a quantity $x$ that
	occurs with probability $p(t)dt$ in the time interval $[t,t+dt)$.  The
	overbar, by contrast, refers to the simple average by time integration over
	one driving period as in Eq.~\eqref{app:Stau}, i.e.\ the phase-average.
	Bayes theorem now tells us that $p_\ell(t+\tau|t) p(t) = p_\ell(t+\tau,t)$,
	which is the joint probability density for events at time $t$ and $t+\tau$
	with $\ell$ events in between.  Summation over $\ell$ removes the latter
	restriction leading to $\sum_\ell p_\ell(t+\tau,t) = p(t+\tau,t)$.  Hence, the
	right-hand side of Eq.~\eqref{app:wtau} can be identified as the term
	required in Eq.~\eqref{app:Sbasic}, namely
	\begin{equation}
		\overline{p(t+\tau,t)}^t = \sum_{\ell=0}^\infty w_\ell(\tau) \equiv w(\tau).
	\end{equation}
	A typical example of $w(\tau)$ for our experiment is shown in
	Fig.~\ref{fig:wtau}.
	
	\begin{figure}
		\centerline{%
		\includegraphics{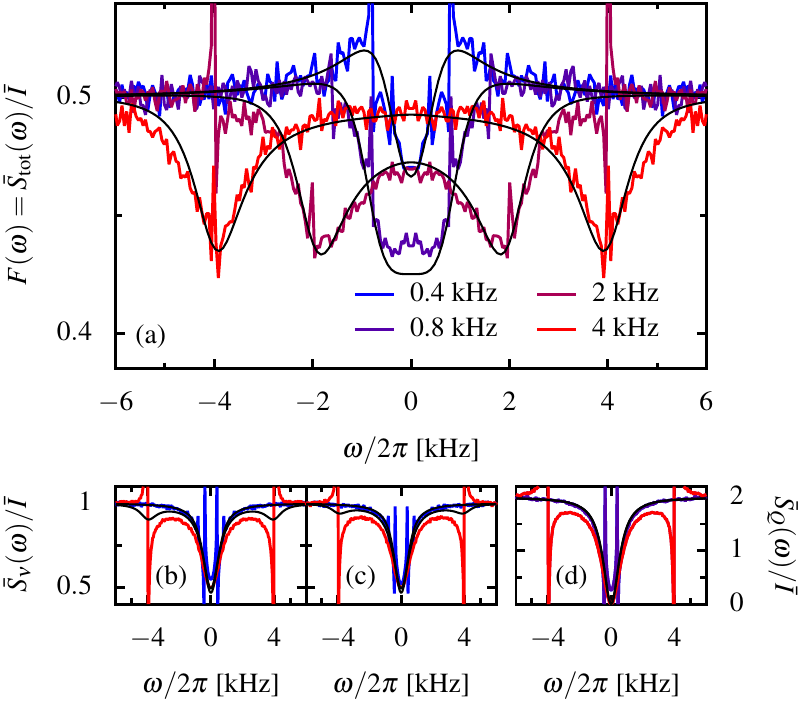}}
		\caption{Data shown in Fig.~\ref{fig:ExpTheo} of the main text, but without
		removing the contribution of the poles caused by $\varphi(\omega)$ at multiples of
		the driving frequencies.}
		\label{fig:FFw_raw}
	\end{figure}%
	
	As $w(\tau)$ is a probability density of events, despite some conditions at
	an earlier time, in the long-time limit, at least on average, it must be
	equal to the mean event rate $\bar\gamma$.  This implies that the
	$\tau$-averaged long-time limit of Eq.~\eqref{app:Sbasic} becomes $0 =
	\lambda\bar\gamma -\bar\gamma^2$ [recall that $\varphi(\tau)$ has zero
	mean], which provides the missing proportionality factor
	$\lambda = \bar\gamma$.
	
	So far, we have assumed $\tau>0$.  To obtain a relation valid for any
	$\tau$, we recall that $S(t,t') = S(t',t)$ is symmetric by definition and
	at $t=t'$ must have a $\delta$-peak that reflects shot noise,
	$\gamma(t)\delta(t-t')$ \cite{Cox1962a, vanKampen1992a, Korotkov1994a}.
	Hence, we find that the auto correlation of the spike train and the sum of
	the waiting time distributions relate as
	\begin{equation}
		\bar S(\tau) + \bar\gamma^2
		= \bar\gamma\delta(\tau) + \bar\gamma w(|\tau|) - \varphi(|\tau|) .
		\label{app:St-wt}
	\end{equation}
	A final Fourier transformation provides the power spectral density
	\begin{equation}
		\bar S(\omega) = \bar\gamma + 2\bar\gamma \int_0^\infty d\tau\,
		\cos(\omega\tau) w(\tau)
		- \sum_k \varphi_k \delta(\omega-k\Omega)
		\label{app:Sw-wt}
	\end{equation}
	with some irrelevant coefficients $\varphi_k$, where $\varphi_0$ absorbs the
	term $\bar\gamma^2$.  This relation forms the basis of our evaluation
	of the experimental data.
	
	For a finite set of sampled data, $w(\tau)$ can be determined only
	within a finite time window.  Then instead of $\delta$-peaks, the discrete
	Fourier transform of $\varphi$ leads to poles of the type
	$e^{i(\omega-k\Omega)\tau_\text{max}} / (\omega-k\Omega)$, as can be seen
	in Fig.~\ref{fig:FFw_raw}.  Their
	contribution can be determined by analyzing the long-time behavior
	of $w(\tau)$ which is $T$-periodic and equal to $\varphi(\tau)$.
	Alternatively, one may compute the discrete Fourier transformation of
	$w(\tau)$ and determine the emerging poles in $\bar S(\omega)+\varphi(\omega)$ by
	fitting the result in small regions around all relevant resonances
	$k\Omega$ to functions $A_k/(\omega-B_k)$ and subtract the result to obtain
	$\bar S(\omega)$.  Owing to the discrete representation of the spectrum and
	to numerical noise, it is advantageous to treat $B_k$ as a fit parameter,
	despite that it is known to match $k\Omega$.  For our data, this approach
	worked rather reliably.
	
	Finally, we have to relate the spike trains with the currents discussed in
	the main text.  To this end, we consider the spike trains $j_L$, $j_R$, and
	$\tilde\jmath$ for the events at the source, at the drain, and at both
	contacts, respectively.  We evaluate each as described above to
	obtain the power spectra $S_{L}$, $S_{R}$, $\tilde S$.  As $\tilde\jmath
	= j_L+j_R$, this also provides the cross correlation $\langle
	j_L,j_R\rangle = \tilde S -S_{L}-S_{R}$.  The latter allows us to obtain
	the power spectra of the displacement current and the total (or
	Ramo-Shockley) current
	\begin{align}
S_{Q} ={}& 2 (S_{L} + S_{R}) - \tilde S,
\\
S_\mathrm{tot} ={}& \kappa_L\kappa_R\tilde S
    + \kappa_L(\kappa_L-\kappa_R)S_{L} + \kappa_R(\kappa_R-\kappa_L)S_{R} ,
\label{app:S}
	\end{align}
	respectively, where one has to pay attention to the sign convention.  For
	the present symmetric case, $\kappa_L=\kappa_R=1/2$, Eq.~\eqref{app:S}
	becomes $S_\mathrm{tot} = \tilde S/4$.  Since the mean spike rate for $\tilde\jmath$ is
	twice the electron current from source to drain, the Fano factor of
	the total current is $F = \tilde F/2$.

\end{document}